\documentclass[preprint,showpacs,preprintnumbers,amsmath,amssymb]{revtex4}

\usepackage{graphicx}
\usepackage{dcolumn}
\usepackage{bm}

\begin{document}

\title{Role of the density dependent symmetry energy in nuclear stopping.\\}

\author{Karan Singh Vinayak}
\author{Suneel Kumar}%
\email{suneel.kumar@thapar.edu}

\affiliation{%
School of Physics and Materials Science, Thapar University, Patiala-147004, Punjab (India)\\
}%

\date{\today}





\maketitle
\section{Introduction}

Information about the nuclear matter under the extreme conditions of temperature and density and the role of symmetry energy under these conditions is still a topic of crucial importance in the present day nuclear physics research. The multifragmentation, collective flow and the nuclear stopping is among the various rare phenomenon which can be observed in heavy-ion collisions at intermediate energies. The nuclear stopping has gained a lot of interest because it provides the possibility to examine the degree of thermalization or equilibration in the matter.\\
 
The symmetry energy which is the difference of the energy per nucleon between pure neutron  matter and symmetric nuclear matter is taken as 32 MeV corresponding to normal nuclear matter density i.e. $\rho = 0.16 fm^{-3}$. This understanding does not remain valid as one goes away from the normal nuclear matter density and symmetric nuclear matter. The equation below gives us the most extensively used theoretical parametrization of how the symmetry energy varies against $\rho$\cite{bao}.

\begin{eqnarray}
E(\rho)=E(\rho_o)(\rho/\rho_o)^{\gamma}
\end{eqnarray}

The larger(smaller) values of the constant $\gamma$ corresponds to stiff(soft) density dependence of the symmetry energy\cite{bao}. Aim of the present study is to pin down the nuclear stopping for different forms of density dependent symmetry energy.\\

\section{ISOSPIN-dependent QUANTUM MOLECULAR DYNAMICS (IQMD) MODEL}

Our calculations are carried out within the framework of Isospin dependent Quantum Molecular Dynamics (IQMD) model. The IQMD\cite{model} is a semi-classical model
which describes the heavy-ion collisions on an event by event basis. For more details, see ref.\cite{model}.\\
In IQMD model, the centroid of each nucleon propagates under the classical equations of motion.
\begin{equation}
\frac{d\vec{r_i}}{dt}~=~\frac{d\it{H}}{d\vec{p_i}}~~;~~\frac{d\vec{p_i}}{dt}~=~-\frac{d\it{H}}
{d\vec{r_i}}~~\cdot
\end{equation}
The $H$ referring to the Hamiltonian reads as:
\begin{equation}
~H~=~\sum_{i}\frac{p_i^2}{2m_i}~+~V^{ij}_{Yukawa}~~+~V^{ij}_{Coul}~~+~V^{ij}_{skyrme}+
~V^{ij}_{symm}\cdot
\end{equation}\\

\section{Preliminary Results}

For the present analysis simulations are carried out for several  thousand events at the incident energy of 100 MeV/nucleon for the systems ${{Ca}^{40}}$ + ${{Ca}^{40}}$ and ${{Ca}^{60}}$ + ${{Ca}^{60}}$  having different isostopic compositions i.e. N/Z=1 and N/Z=1.5 respectively. The whole analysis is performed for global nuclear stopping. One can use the anisotropy ratio(R) as a probe for the degree of stopping\cite{stop}.\\
\begin{figure}
\hspace{-2.0cm}\includegraphics[scale=0.30]{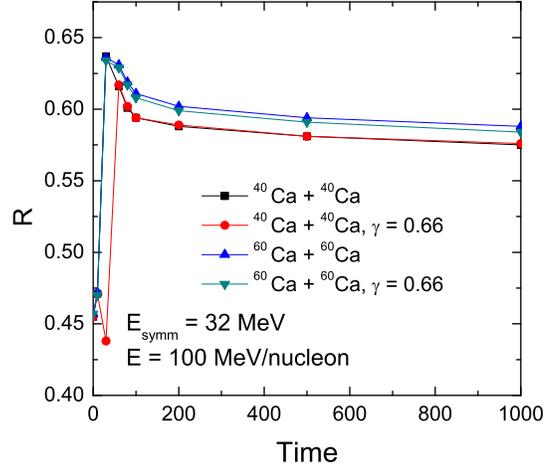}
\caption{\label{fig:1} Time evolution of the nuclear stopping(anisotropy ratio) for different isotopes of calcium}.
\end{figure}

\begin{equation}
~R~=~\frac{2}{\pi}~\frac{[~\sum_{i}p_{\bot}(i)~]}{[~\sum_{i}p_{||}(i)~]}\cdot
\end{equation}\\

where the summation runs over all nucleons. The transverse and longitudinal momenta are $ p_{\bot}=\sqrt{p_{x}^{2}(i)+p_{y}^{2}(i)}~and~ p_{||}(i)=p_{z}(i)$ respectively.

In fig.1, we display the time evolution of the anisotropy ratio for the above said systems having different isotopic compositions, for the symmetry energy of 32 MeV and $32(\rho/\rho_o)^{\gamma}$ MeV for ${\gamma}=0.66$ to study the effect of symmetry energy on global nuclear stopping. Our trends show the large sensitivity of the symmetry energy for the nuclear stopping. The anisotropy ratio tends to increase by 10 percent for the system having large isotopic asymmetry as compared to symmetric system. The various forms of the symmetry energy does not affect the symmetric systems. The mild effect is observed for the asymmetric system with ${\gamma}=0.66$ as compared to full symmetry energy
strength. Which concludes that the role of symmetry energy tends to increase as one moves away from the symmetric nuclear matter. The different forms of the symmetry energy affects nuclear stopping of asymmetric systems, wheres as no such affect is seen in case of symmetric systems.\\ 

\begin{figure}
\hspace{-2.0cm}\includegraphics[scale=0.30]{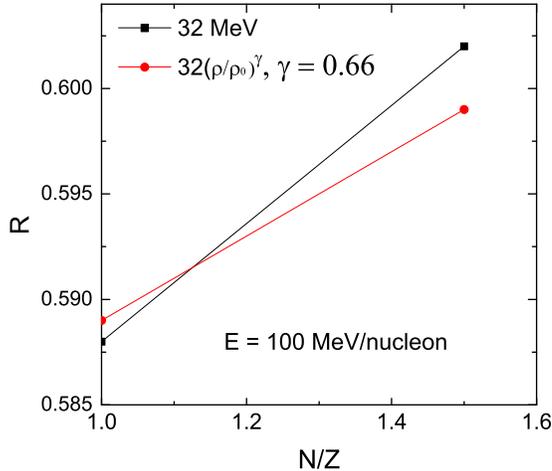}
\caption{\label{fig:2} The N/Z dependence of the nuclear stopping(anisotropy ratio).}
\end{figure}

In fig.2, we display the anisotropy ratio with respect to the N/Z ratio for the symmetry energy of 32 MeV and $32(\rho/\rho_o)^{\gamma}$ MeV for ${\gamma}=0.66$. For the symmetric system the ${\gamma}=0.66$ produces more stopping as compared to  full symmetry energy strength, i.e. 32 MeV. In case of the asymmetric systems difference of the nuclear stopping is more as compared to the symmetric systems for the different forms of the symmetry energy. Therefore, the symmetry energy tends to play larger role for asymmetric systems as concluded above in fig.1. However, more stopping is observed for the constant form of the symmetry energy.\\



\end{document}